\documentclass[aps,prd,twocolumn,amsfonts,amsmath,showpacs]{revtex4}
\usepackage{amsmath}
\usepackage[dvips]{graphicx}
\begin{document}
\title{CMB Angular Power Spectrum from Correlated Primordial Fluctuation}
\author{B.Yu}
\email{yubo@chenwang.nju.edu.cn}
\affiliation{Department of Physics, Nanjing University, Nanjing 210093, China}
\author{T.Lu}
\email{t.lu@mail.pmo.ac.cn}
\affiliation{Purple Mountain Observatory, Chinese Academy of Sciences,
Nanjing 210008, China}
\affiliation{Joint Center for Particle, Nuclear Physics and Cosmology, Nanjing University-Purple Mountain Observatory, Nanjing 210093, China}
\date{\today.}

\begin{abstract}
The usual inflationary scenario predicts a Gaussian random primordial density fluctuation, different Fourier modes of which do not correlate with each other. In this paper we propose a correlation between these different modes. A simple case is that these different Fourier modes correlate with each other following a Gaussian function. For such a primordial density fluctuation we calculate the CMB angular power spectrum and find that its amplitude decreases but the decrease is different for different $l$. This feature can be used to constrain the the correlation strength from the real data.
\end{abstract}
\maketitle

\section{Introduction}
In the inflationary scenario, the primordial fluctuation comes from the quantum vacuum fluctuation of the inflaton which drive the inflation \cite{Mukhanov:1981xt,Starobinsky:1982ee,Guth:1982ec,Hawking:1982cz}. Then the quantum fluctuation becomes classical when the corresponding scale goes out of the horizon. The primordial fluctuation produced in this method is regarded to be Gaussian random and scale invariant, which are confirmed by many astronomical observations \cite{Smoot:1992td,Bennett:1996ce,Spergel:2006hy,Hinshaw:2006ia}.

This explanation of the origin of the primordial fluctuation seems reasonable and successful. However it is not a rigorous theory and needs more experimental tests. Besides the inflation, there exist other ideas to produce the primordial fluctuations, for example, the string gas model \cite{Nayeri:2005ck,Brandenberger:2006xi,Brandenberger:2006vv,Brandenberger:2006pr}. Even in the inflationary scenario, there are many kinds of modification to the primordial power spectrum, for example, the case that translational invariance is broken \cite{Carroll:2008br} or rotational invariance is broken \cite{Ackerman:2007nb,Pullen:2007tu,Groeneboom:2008fz,ArmendarizPicon:2008yr}. The quantum-classical transition in the inflationary scenario is also criticized by \cite{Perez:2005gh}. Considering these uncertainties, our paper is motivated by checking one of the main features of the Gaussian random fluctuation from the experiments: the different modes of the primordial fluctuation in the Fourier space are independent from each other. As we know, the Cosmic Microwave Background (CMB) is the relics of early universe, which can provide precise measurement on the primordial fluctuation. So in this paper we introduce a small correlation between different modes of the primordial fluctuation (in Fourier space) and then study how the correlation affects the CMB angular power spectrum.

This paper is organized as follows. In section II we introduce the above mentioned correlations of the primordial fluctuation in Fourier space. In section III we calculate the CMB angular power spectrum. The numerical results are presented in section IV and the discussions are given in section V.
\section{Primordial fluctuation}
The meaning of the Gaussian random primordial density fluctuation is $< \xi(\vec{k}) \xi^{*}(\vec{k}^{'}) > = P^{s}(k) \delta^{3}(\vec{k}-\vec{k}^{'})$, where $\xi(\vec{k})$ is the Fourier mode of the primordial density fluctuation $\xi(\vec{x})$. The delta function indicates that different mode does not correlate with each other. One can relax this property by allowing different modes to correlate. Generally speaking, one can rewrite the above equation as $< \xi(\vec{k}) \xi^{*}(\vec{k}^{'}) > = f(\vec{k},\vec{k}^{'})$, where $f(\vec{k},\vec{k}^{'})$ is not required to be a delta function any more. However $f(\vec{k},\vec{k}^{'})$ is not arbitrary, it must satisfy some constraints since $\xi(\vec{x})$ is a real field. $< \xi(\vec{k}) \xi^{*}(\vec{k}^{'}) > ^{*} = < \xi(\vec{k}^{'}) \xi^{*}(\vec{k}) >$ gives $f^{*}(\vec{k},\vec{k}^{'})= f(\vec{k}^{'},\vec{k})$, which means the real part of $f(\vec{k},\vec{k}^{'})$ is symmetric under the interchange of $\vec{k},\vec{k}^{'}$, $Re f(\vec{k},\vec{k}^{'}) = Re f(\vec{k}^{'},\vec{k})$, and the imaginary part of $f(\vec{k},\vec{k}^{'})$ is antisymmetric under the interchange of $\vec{k},\vec{k}^{'}$, $Im f(\vec{k},\vec{k}^{'}) = -Im f(\vec{k}^{'},\vec{k})$. Besides this, $\xi^{*}(\vec{k}) = \xi(-\vec{k})$, which indicates that $< \xi^{*}(-\vec{k}) \xi(-\vec{k}^{'}) > = f(-\vec{k}^{'},-\vec{k}) =f(\vec{k},\vec{k}^{'}) = < \xi(\vec{k}) \xi^{*}(\vec{k}^{'}) >$. Thus one obtains $Re f(-\vec{k},-\vec{k}^{'}) = Re f(-\vec{k}^{'},-\vec{k}) = Re f(\vec{k},\vec{k}^{'})$ and $Im f(-\vec{k},-\vec{k}^{'}) = -Im f(-\vec{k}^{'},-\vec{k}) = -Im f(\vec{k},\vec{k}^{'})$. $P^{s}(k) \delta^{3}(\vec{k}-\vec{k}^{'})$ is such a special function with a zero imaginary part.

However, these constraints provide no positive information on the function $f(\vec{k},\vec{k}^{'})$. We must construct a concrete form of the above function. It is not strange to require that delta function is a good approximation to this function, and noting that when $\sigma \rightarrow 0$, $\frac{1}{\sqrt{2\pi\sigma}}\exp{[-\frac{(x-y)^{2}}{2\sigma}]} \rightarrow \delta (x-y)$, one can use a very simple form
\begin{eqnarray}
(\frac{1}{\sqrt{2\pi\sigma_{0} k^{2}_{o}}})^{3}\exp{[-\frac{(\vec{k}-\vec{k}^{'})^{2}}{2\sigma_{0} k^{2}_{o}}]}
\end{eqnarray}
to replace the delta function. When $\sigma_{0} \rightarrow 0$, it approaches $\delta^{3}(\vec{k}-\vec{k}^{'})$. Here $\sigma_{0}$ is a pure number and $k_{0}$ is the comoving wave number. This is a Gaussian form function, which is completely determined by the parameter $\sigma \equiv \sigma_{0} k^{2}_{0}$ and its maximum is located at $\vec{k} = \vec{k}^{'}$. The FWHM is $2\sqrt{2 \ln 2 \sigma} $. Then in order to satisfy the above constraints, one can generalize $P^{s}(k)$ to $P^{s}(\sqrt{k k^{'}})$ or $P^{s}(\frac{k+k^{'}}{2})$. Their difference can be ignored in the weak correlation case ($\sigma$ is small), but cannot be ignored in the strong correlation case ($\sigma$ is great). However, it is easy to note that the strong correlation case is excluded by the CMB observation, so in this paper we only take
\begin{eqnarray}
f(\vec{k},\vec{k}^{'}) = P^{s}(\frac{k+k^{'}}{2}) (\frac{1}{\sqrt{2\pi\sigma_{0} k^{2}_{o}}})^{3}\exp{[-\frac{(\vec{k}-\vec{k}^{'})^{2}}{2\sigma_{0} k^{2}_{o}}]} 
\label{fkk}
\end{eqnarray}
as an example to study the correlation effects on CMB angular power spectrum. In fact, the following discussions not only apply to primordial fluctuation with form (\ref{fkk}) but also to such a form $f(\vec{k},\vec{k}^{'}) = f(k,k^{'},|\vec{k}-\vec{k}^{'}|)$. It can be rewritten as
\begin{eqnarray}
f(k,k^{'},|\vec{k}-\vec{k}^{'}|) = \int d^{3}\vec{\Delta} f(k,k^{'},|\vec{\Delta}|) \delta^{3} (\vec{k}-\vec{k}^{'} - \vec{\Delta})
\end{eqnarray}
\section{CMB Power Spectrum}
\subsection{Two point correlator}
For scalar type of primordial density fluctuation, the CMB temperature anisotropy at direction $\vec{n}$ can be expressed as \cite{Dodelson:2003ft,Zaldarriaga:1996xe}
\begin{eqnarray}
T^{s}(\vec{n}) & = & \int d^{3}\vec{k} \Delta^{s}_{T}(\eta_{0},k,\hat{k} \cdot \hat{n}) \xi(\vec{k}) \nonumber\\
& = & \int d^{3}\vec{k} \int^{\eta_{0}}_{0} d\eta S_{T}^{s}(\eta,k) e^{-ix\hat{k} \cdot \hat{n}} \xi(\vec{k})
\label{talm}
\end{eqnarray}
where $S_{T}^{s}(\eta,k)$ is the transfer function, $x \equiv k(\eta_{0} - \eta)$, $\eta$ is the conformal time and $\eta_{0}$ is the conformal time of today. Following the standard methods, the temperature anisotropy can be expanded by the spherical harmonic coefficients
\begin{eqnarray}
a_{lm}=\int d\Omega Y_{lm}^{*}(\vec{n}) T^{s}(\vec{n}).
\end{eqnarray}
An important property of the modified primordial power spectrum $< \xi(\vec{k}) \xi^{*}(\vec{k}^{'}) > = f(k,k^{'},|\vec{k}-\vec{k}^{'}|)$ is that the CMB anisotropy produced by such a form of primordial density fluctuation is statistically rotational invariant which means that for any rotation $R$ one always have
\begin{eqnarray}
< T^{s}(R \vec{n}^{'}) T^{s}(R \vec{n}) > = < T^{s}(\vec{n}^{'}) T^{s}(\vec{n}) >.
\label{rt}
\end{eqnarray}
The definite proof of Eq.(\ref{rt}) is postponed to Appendix. With this property, one can obtain the following conclusion \cite{Hu:2001fa}
\begin{eqnarray}
< a_{lm} a^{*}_{l^{'}m^{'}} > = \delta_{ll^{'}} \delta_{mm^{'}} C_{l}.
\end{eqnarray}
So the angular power spectrum $C_{l}$ is sufficient to reflect all the two point correlators of CMB anisotropy.
\subsection{Power spectrum}
The CMB angular power spectrum can be calculated by
\begin{eqnarray}
C_{l} & = & \dfrac{1}{2l+1} \sum_{m} < a^{*}_{lm} a_{lm} >  \nonumber\\
& = & \dfrac{1}{2l+1} \sum_{m} < (\int d\Omega Y_{lm}^{*}(\vec{n}^{'}) T(\vec{n}^{'}))^{*} \nonumber\\
&& (\int d\Omega Y_{lm}^{*}(\vec{n}) T(\vec{n})) >
\label{ctl}
\end{eqnarray}
Substituting Eq.(\ref{talm}) into Eq.(\ref{ctl}), one obtains
\begin{eqnarray}
C_{l}  = && \dfrac{(4\pi)^{2}}{2l+1} \sum_{m} \int d^{3}\vec{k}^{'} d^{3}\vec{k} G_{l}(k^{'}) G_{l}(k)  \nonumber\\
&&Y_{lm}(\hat{k}^{'}) Y_{lm}^{*}(\hat{k}) < \xi(\vec{k}) \xi^{*}(\vec{k}^{'}) > \nonumber\\
= && \int d^{3}\vec{\Delta} F(\vec{\Delta}), 
\end{eqnarray}
where the formula ${\rm e}^{i \vec{k} \cdotp \vec{x}} = 4 \pi \sum_{lm} i^l j_l(kx)Y^*_{lm}(\hat{k}) Y_{lm}(\hat{x})$ has been used and $F(\vec{\Delta})$ is defined by
\begin{eqnarray}
F(\vec{\Delta}) & \equiv & \dfrac{(4\pi)^{2}}{2l+1} \sum_{m} \int d^{3}\vec{k}^{'} d^{3}\vec{k} G_{l}(k^{'}) G_{l}(k) Y_{lm}(\hat{k}^{'})  \nonumber\\
&& Y_{lm}^{*}(\hat{k}) f(k,k^{'},|\vec{\Delta}|) \delta^{3} (\vec{k}-\vec{k}^{'} - \vec{\Delta}),
\label{fd}
\end{eqnarray}
where $G_{l}(k) \equiv \int^{\eta_{0}}_{0} d\eta S_{T}^{s}(\eta,k) j_{l}(x)$. In the Appendix we have proved
\begin{eqnarray}
F(R \vec{\Delta}) = F(\vec{\Delta}).
\end{eqnarray}
So, $F(\vec{\Delta})$ is only a function of $\Delta \equiv |\vec{\Delta}|$. Thus
\begin{eqnarray}
C_{l} = 4\pi \int^{\infty}_{0} \Delta^{2} d\Delta F(\Delta).
\label{cf}
\end{eqnarray}
Introducing new variables $\vec{p} = \vec{k} + \vec{k}^{'}, \vec{p} = \vec{k} - \vec{k}^{'}$, and $|\dfrac{\partial(\vec{k},\vec{k}^{'})}{\partial(\vec{p},\vec{p}^{'})}| = \frac{1}{8}$, one obtains
\begin{eqnarray}
F(\vec{\Delta}) &=& \dfrac{(4\pi)^{2}}{8(2l+1)} \sum_{m}  \int d^{3}\vec{p} d^{3}\vec{p}^{'}  G_{l}(k^{'}) G_{l}(k) Y_{lm}(\hat{k}^{'})  \nonumber\\
&& Y_{lm}^{*}(\hat{k}) f(k,k^{'},|\vec{\Delta}|) \delta^{3} (\vec{p}^{'} - \vec{\Delta}) \nonumber\\
&=& \dfrac{(4\pi)^{2}}{8(2l+1)} \sum_{m}  \int d^{3}\vec{p} G_{l}(k^{'}) G_{l}(k) Y_{lm}(\hat{k}^{'})  \nonumber\\
&& Y_{lm}^{*}(\hat{k}) f(k,k^{'},\Delta)  \nonumber\\
&=& \dfrac{\pi}{2} \int d^{3}\vec{p} G_{l}(k^{'}) G_{l}(k) P_{l}(\hat{k^{'}} \cdot \hat{k}) f(k,k^{'},\Delta),
\label{fde}
\end{eqnarray}
where $\vec{k}=\dfrac{\vec{p}+\vec{p}^{'}}{2}, \vec{k}=\dfrac{\vec{p}-\vec{p}^{'}}{2}$. Since
$F(\vec{\Delta})$ is only a function of $\Delta$, one can safely set $\vec{\Delta}$ in the $z$ direction $\vec{\Delta} = \Delta \vec{e_{z}}$, then $k = \dfrac{\sqrt{p_{x}^{2}+p_{y}^{2}+(p_{z}+\Delta)^{2}}}{2}$, $k^{'} = \dfrac{\sqrt{p_{x}^{2}+p_{y}^{2}+(p_{z}-\Delta)^{2}}}{2}$, $\hat{k^{'}} \cdot \hat{k} =\dfrac{p_{x}^{2}+p_{y}^{2}+p_{z}^{2}-\Delta^{2}}{4kk^{'}}$. Noting that the integrand is only a function of $p_{x}^{2}+p_{y}^{2}$, and using
\begin{eqnarray}
&&\int_{-\infty}^{+\infty} dp_{x} \int_{-\infty}^{+\infty} dp_{y} f(p_{x}^{2}+p_{y}^{2})  \nonumber\\
& = & \int^{\infty}_{0} \bar{p} d\bar{p} \int^{\pi}_{0} d\theta f(\bar{p}^{2})  \nonumber\\ 
& = & 2\pi \int^{\infty}_{0} \bar{p} d\bar{p} f(\bar{p}^{2}),
\end{eqnarray}
one obtains
\begin{eqnarray}
C_{l} = & & 4\pi^{3} \int^{\infty}_{0} \Delta^{2} d\Delta \int^{+\infty}_{-\infty} dp_{z} \int^{\infty}_{0} \bar{p} d\bar{p} G_{l}(k^{'}) G_{l}(k) \nonumber\\
& & P_{l}(\hat{k^{'}} \cdot \hat{k}) f(k,k^{'},\Delta).
\end{eqnarray}
Finally noting that the integrand is an even function of $p_{z}$, one obtains a simple form,
\begin{eqnarray}
C_{l} & = & \frac{\pi}{2} (4\pi)^{2} \int^{\infty}_{0} \Delta^{2} d\Delta \int^{\infty}_{0} dp_{z} \int^{\infty}_{0} \bar{p} d\bar{p}  \nonumber\\
& & G_{l}(k^{'}) G_{l}(k) P_{l}(\hat{k^{'}} \cdot \hat{k}) f(k,k^{'},\Delta),
\label{ctlf}
\end{eqnarray}
where $G_{l}(k) \equiv \int^{\eta_{0}}_{0} d\eta S_{T}^{s}(\eta,k) j_{l}(x)$, $k = \dfrac{\sqrt{\bar{p}^{2}+(p_{z}+\Delta)^{2}}}{2}$, $k^{'} = \dfrac{\sqrt{\bar{p}^{2}+(p_{z}-\Delta)^{2}}}{2}$ and $\hat{k^{'}} \cdot \hat{k} =\dfrac{\bar{p}^{2}+p_{z}^{2}-\Delta^{2}}{4kk^{'}}$.
\section{Numerical Results}
In this paper we use the Boltzmann code CMBFAST (\cite{Seljak:1996is}) to do numerical calculation and adopt a flat $\Lambda CDM$ model with cosmological parameters: $\Omega_b = 0.045$, $\Omega_c = 0.225$, $\Omega_{\Lambda} = 0.73$, $H_0 = 70$, $\tau = 0.08$, $n_{s} = 1$ and $\Delta^2_{\xi}=1$. Where $\tau$ is the optical depth, $n_{s}$ is spectral index and $\Delta_{\xi}$ is the primordial scalar perturbation amplitude. In CMB power spectrum calculation $\Delta^2_{\xi}$ is only a normalization factor, the value of which is determined by observational data. Thus we can safely set it to be unit (usually the value of $\Delta^2_{\xi}$ is taken to be about $2.35 \times 10^{-10}$), which will not affect the following discussion, since we mainly care about the relative change of the power spectrum.

\begin{figure}
\begin{center}
\includegraphics[width=0.4\textwidth]{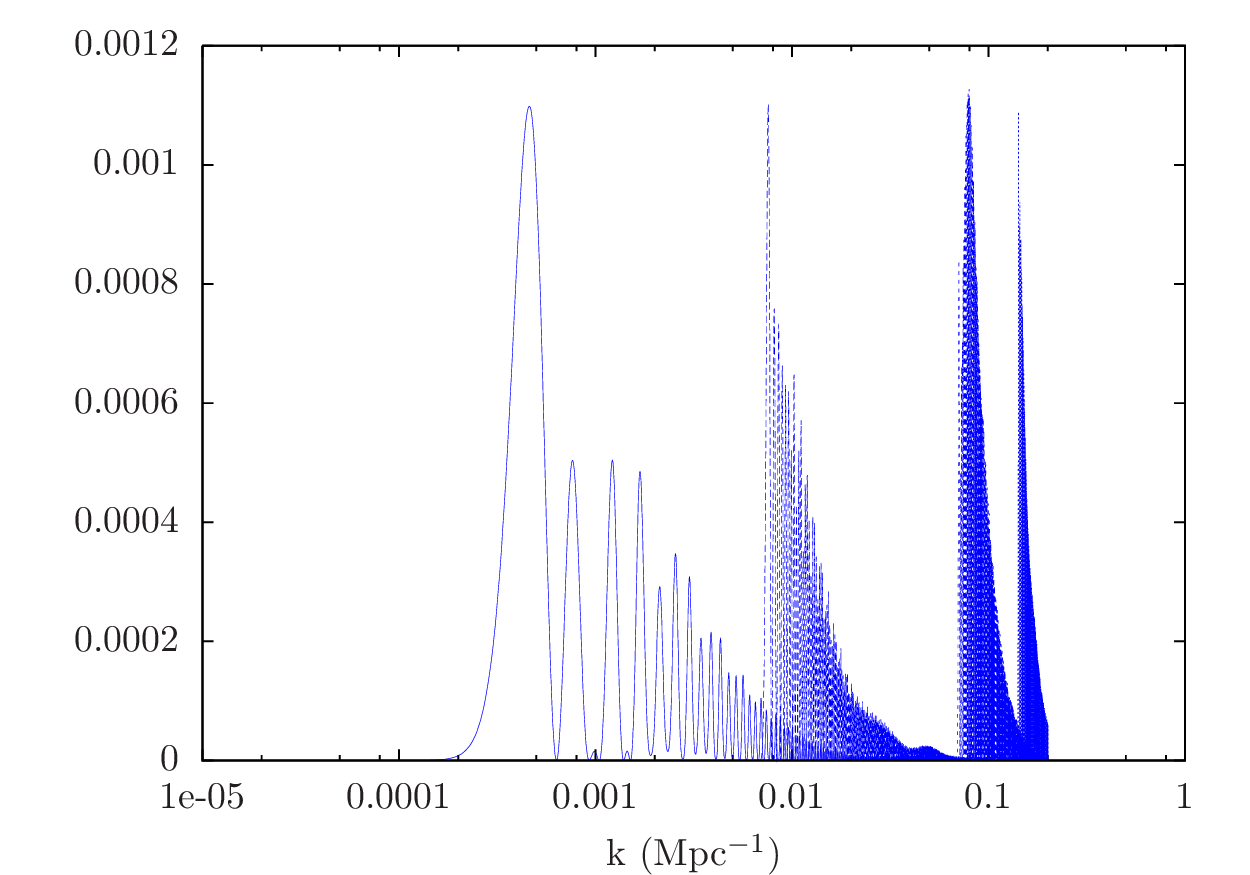}
\end{center}
\caption{Dependence of $(G_{l}(k))^{2}$ on $k$, from left to right these curves correspond to $(G_{5}(k))^{2},34*(G_{100}(k))^{2},5500*(G_{1000}(k))^{2},37000*(G_{2000}(k))^{2}$ respectively.}
\label{f1}
\end{figure}

\begin{figure}
\begin{center}
\includegraphics[width=0.4\textwidth]{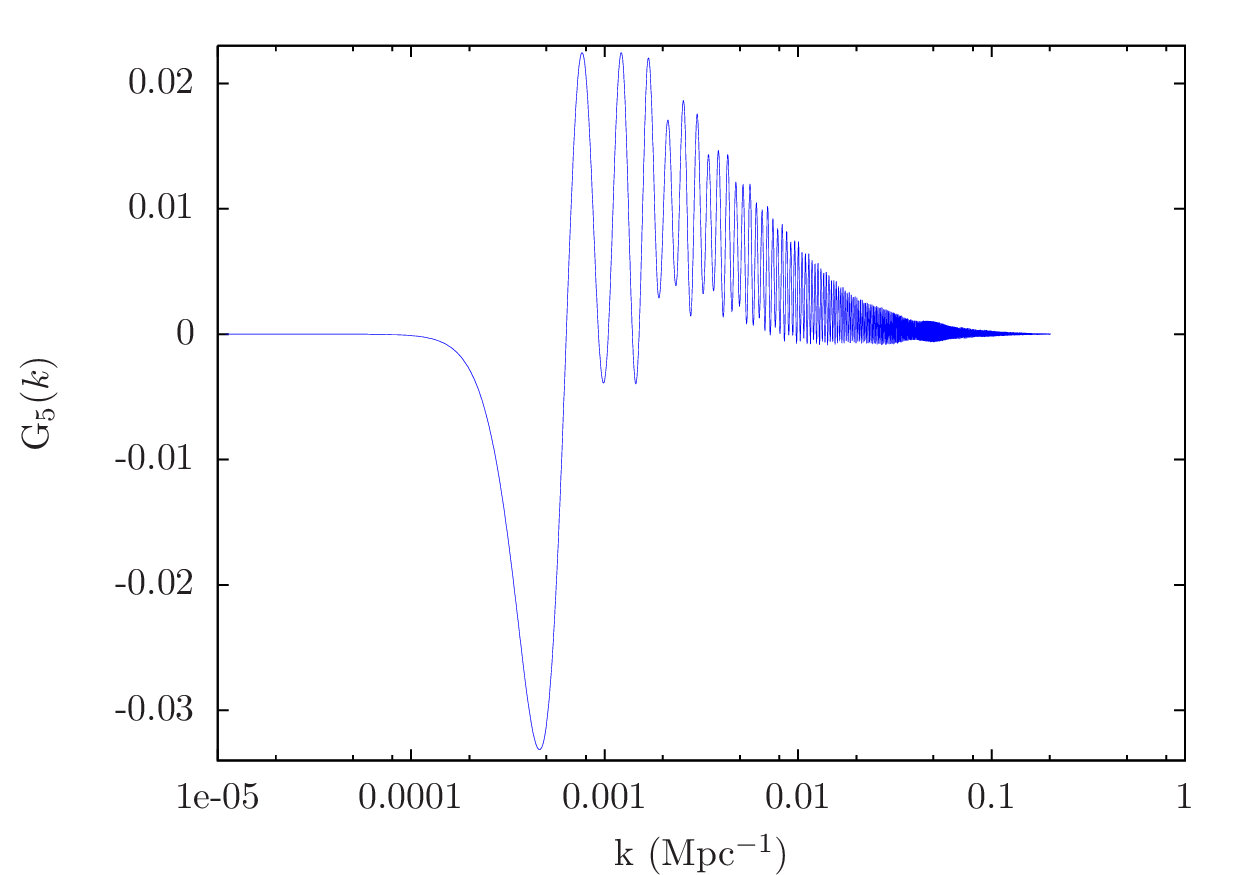}
\end{center}
\caption{Dependence of $G_{5}(k)$ on $k$.}
\label{f2}
\end{figure}

\begin{figure}
\begin{center}
\includegraphics[width=0.4\textwidth]{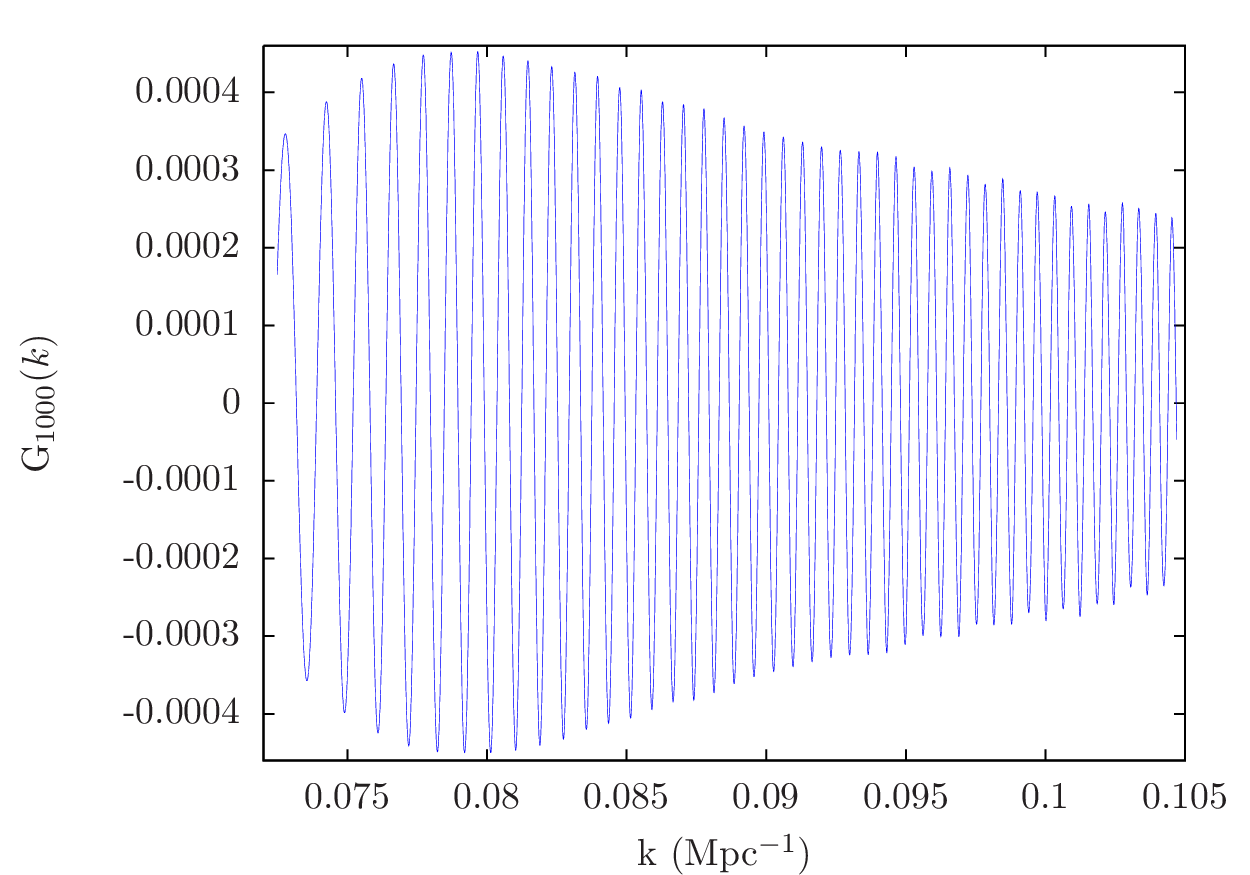}
\end{center}
\caption{Dependence of $G_{1000}(k)$ on $k$.}
\label{f3}
\end{figure}

\begin{figure}
\begin{center}
\includegraphics[width=0.4\textwidth]{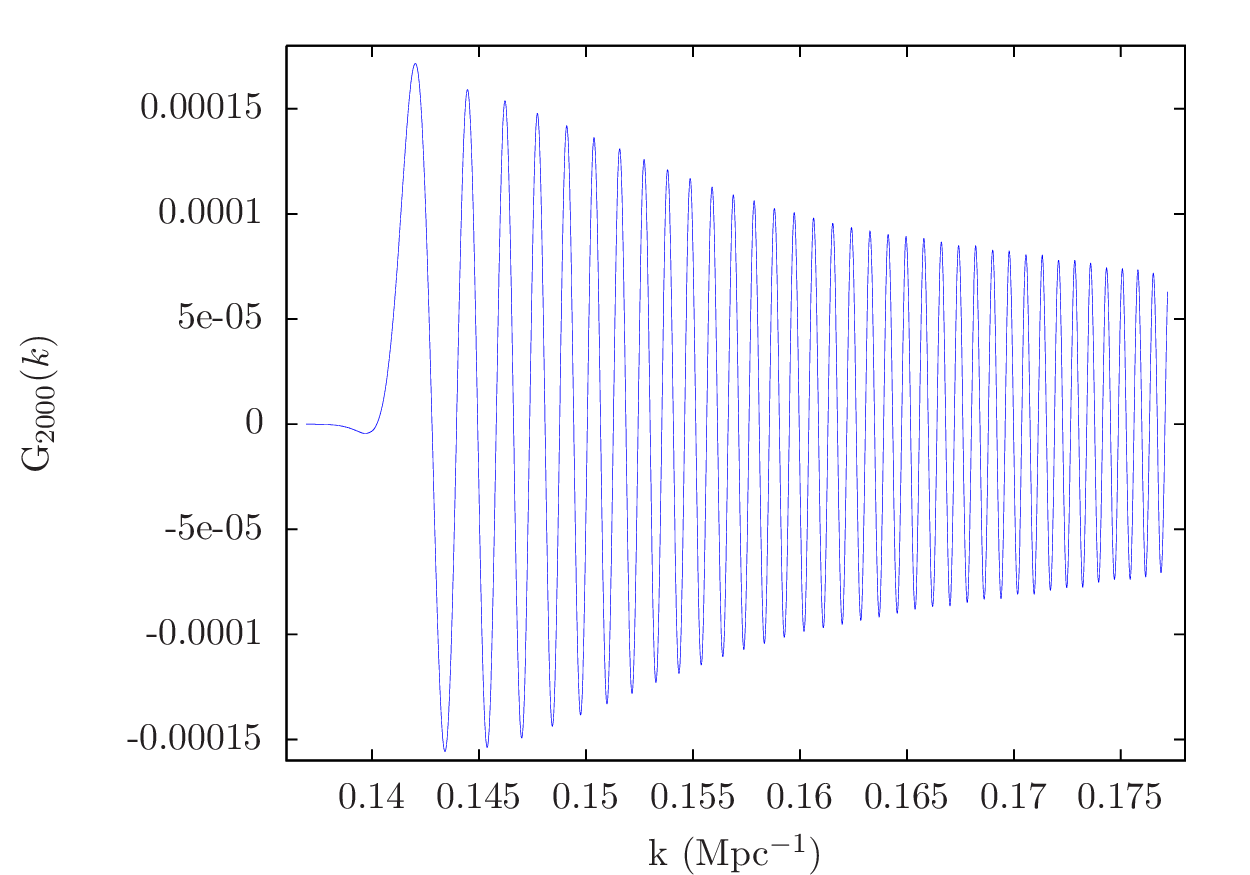}
\end{center}
\caption{Dependence of $G_{2000}(k)$ on $k$.}
\label{f4}
\end{figure}

First we calculate $(G_{l}(k))^{2} = (\int^{\eta_{0}}_{0} d\eta S_{T}^{s}(\eta,k) j_{l}(x))^{2}$ for $l=5,100,1000,2000$ as a function of $k$. When there is no correlation between the different modes, the CMB angular power spectrum is $C_{l} = (4\pi)^{2} \int_{0}^{\infty} k^{2}dk P^{s}(k) (G_{l}(k))^{2}$, which reduces to $C_{l} = (4\pi)^{2} \Delta^2_{\xi} \int_{0}^{\infty} d\ln k (G_{l}(k))^{2}$ when the spectrum is scale invariant. From Fig.(\ref{f1}) it is easy to note that for different $l$ the main contribution to $C_{l}$ comes from different region of $k$. The larger is $l$, the main contributions to $C_{l}$ comes from larger $k$.
Then we plot $G_{l}(k) = \int^{\eta_{0}}_{0} d\eta S_{T}^{s}(\eta,k) j_{l}(x)$ for $l=5,1000,2000$ in their main region (see Fig.(\ref{f2},\ref{f3},\ref{f4})). One note that when $l=5$ $G_{l}(k)$ is almost positive except for small $k$  where there exist a negative peak. One also note that for $l=2000$ the average period is longer than that for $l=1000$.

\begin{figure}
\begin{center}
\includegraphics[width=0.4\textwidth]{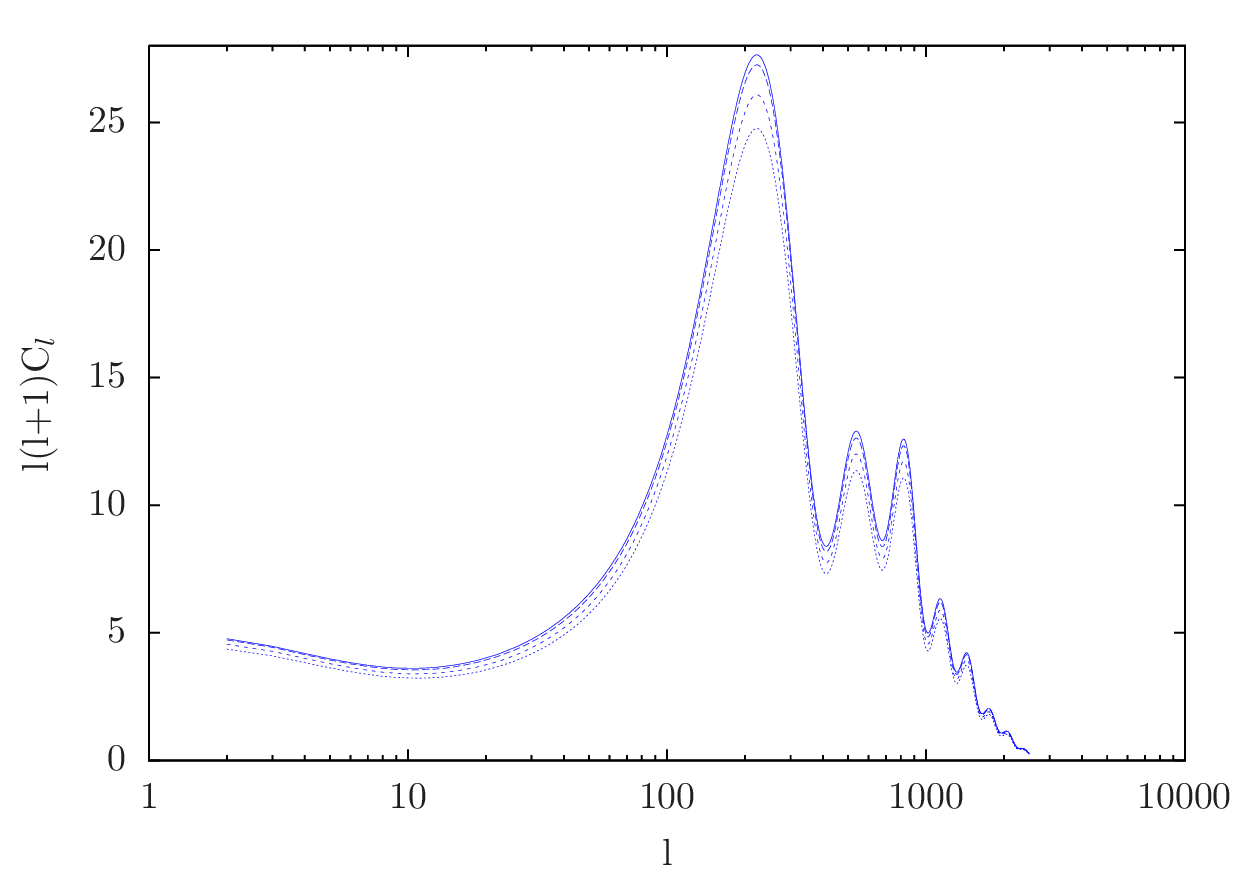}
\end{center}
\caption{Unnormalized CMB angular power spectrum $C_{l}$ as a function of $l$. The solid line corresponds to the Gaussian random case, while the long dashed line, the short dashed line and the dotted line corresponds to $\sigma = 10^{-10},5*10^{-10}$ and $10^{-9}$ respectively.}
\label{power}
\end{figure}

\begin{figure}
\begin{center}
\includegraphics[width=0.4\textwidth]{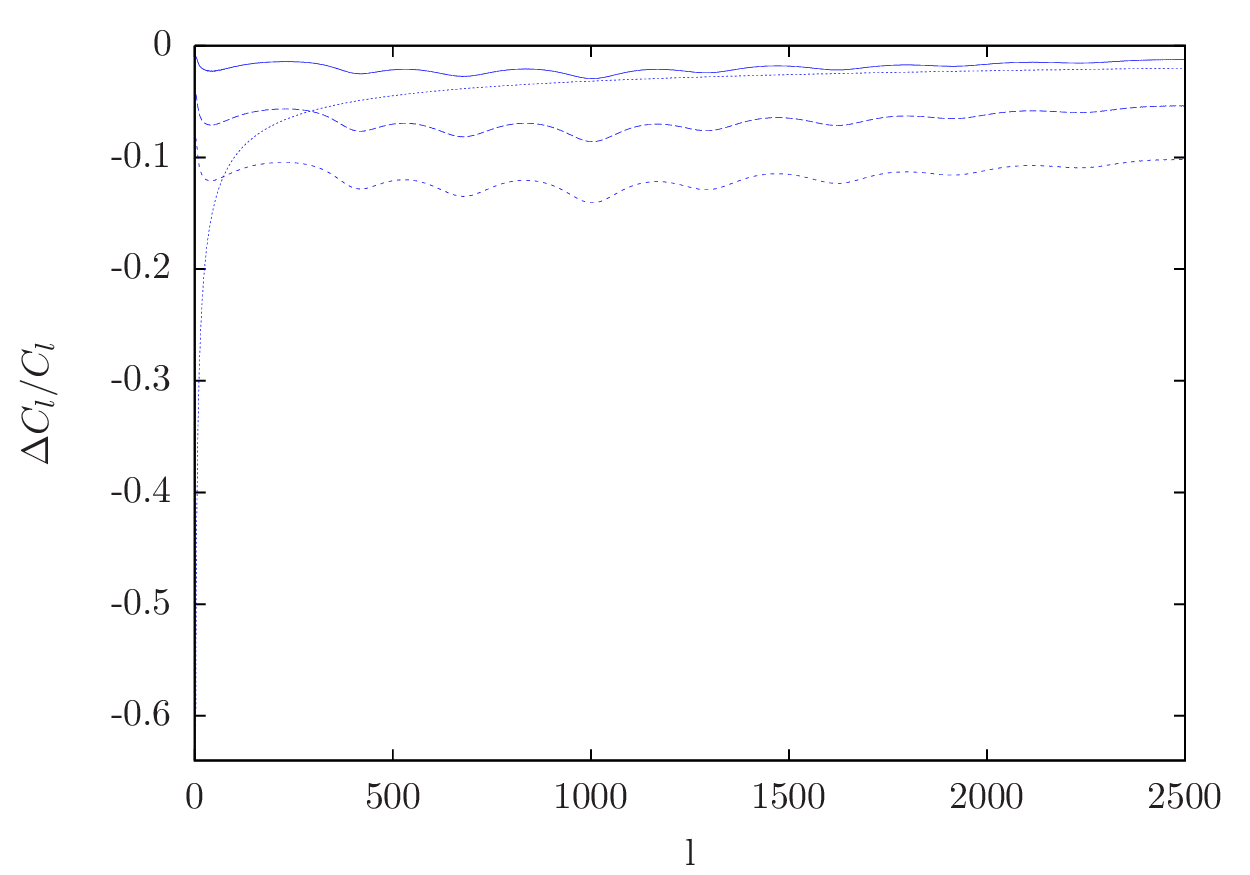}
\end{center}
\caption{Dependence of $\frac{\Delta C_{l}}{C_{l}}$ on $l$. $\Delta C_{l}$ is defined as the CMB angular power spectrum with correlation minus the one without correlation. The solid line, long dashed line and short dashed line correspond to $\sigma = 10^{-10},5*10^{-10}$ and $10^{-9}$ respectively. The dotted line is the cosmic variance of the angular power spectrum $\Delta C_{l} / C_{l} = \sqrt{2/(2l+1)}$ in the uncorrelated case (here $-\sqrt{2/(2l+1)}$ is plotted).}
\label{change}
\end{figure}

The CMB angular power spectra are plotted in Fig.(\ref{power}) and the relative angular power spectra changes and the relative cosmic variance are plotted in Fig.(\ref{change}). It is easy to note that the amplitude of angular power spectrum decreases when correlation is introduced. The greater is $\sigma$, the lower is the amplitude. From Fig.(\ref{rt}) it can be seen that the angular power spectrum decreases more significantly at the neighborhood of $l=1000$ than at smaller or larger $l$. This is because of the phenomenon we mentioned in the last paragraph. Numerical calculation tells us that the contribution of $F(\Delta)$ to $C_{l}$ (Eq.(\ref{cf})) concentrate at the neighbourhood of $\Delta = \sqrt{2 \sigma}$ and $\hat{k^{'}} \cdot \hat{k} \approx 1$ in this case when $\sigma$ is small (so $P_{l}(\hat{k^{'}} \cdot \hat{k}) \approx 1$). Thus one can simply consider such a quantity $\tilde{C}_{l} \equiv (4\pi)^{2}\int_{0}^{\infty} d \ln k G_{l}(k) * G_{l}(k + \delta k)$ (refer to Eq.(\ref{fde})) and compare it with $C_{l} = (4\pi)^{2} \int_{0}^{\infty} d\ln k (G_{l}(k))^{2}$  to understand the above mentioned phenomenon, where $\delta k$ is a small fixed value. From Fig.(\ref{f2},\ref{f3},\ref{f4}) one know that $G_{l}(k)$ oscillate with $k$, so one can further consider such a function,
\begin{displaymath}
f(x) = \left\{  \begin{array}{cc}
0  &  x<-1 \\
x+1  &  -1 \leq x \leq 0 \\
-x+1  &  0 \leq x \leq 1 \\
0 & x>1.
\end{array} \right.
\end{displaymath}
The quantity $ D_{1} \equiv \int f^{2}(x) dx = \frac{2}{3}$, while another quantity $D_{2} (\delta) \equiv \int f(x)f(x + \delta) dx = \frac{2}{3} - \delta^{2}(1-\frac{\delta}{2})$ (here $\delta \ll 1$). One can see that $D_{2} < D_{1}$. As the same reason, the angular power spectrum decrease when the primordial density fluctuation is correlated in Fourier space. At the same time, when $ 0 < \delta_{1} < \delta_{2}$, $D_{2} (\delta_{1}) > D_{2} (\delta_{2})$. Due to this, for a fix $\delta k$, $\tilde{C}_{l}$ decrease less in the case that $G_{l}(k)$ oscillates with a longer average period than the decrease in the case that $G_{l}(k)$ oscillates with a shorter average period. Thus the angular power spectrum at large $l$ decrease less than that at about $l \sim 1000$, since $G_{1000}(k)$ oscillates with a shorter average period compared with $G_{2000}(k)$. At last one note that when $G_{l} (k)$ is always positive, the integrand $G_{l}(k) * G_{l}(k + \delta k)$ is also positive, while when $G_{l} (k)$ is positive and negative alternate, the integrand $G_{l}(k) * G_{l}(k + \delta k)$ is not always positive. So when $G_{l}(k)$ is always positive $\tilde{C}_{l}$ tends to decrease less than the case that $G_{l}(k)$ takes positive value and negative value alternately. This is the reason that the angular power spectrum decrease less at small $l$ than that at $l=1000$. These features can be used to constrain the correlation strength parameter $\sigma$ from the real data.
\section{Discussion}
In this paper we only discussed a special type of the correlation of the primordial density fluctuation in the Fourier space. There exist other forms of correlation. From the point of experimental view, the correlation discussed in our paper is typical because the correlation between two modes is large only when they are almost the same. However, it is possible that the correlation in the Fourier space is scale dependent. Such possibilities can not be excluded. A better method is to divide these possible forms into several types, and then consider their effects on CMB. Besides CMB temperature, the effects on CMB polarization and large scale structure also need a further study.

\begin{acknowledgments}
We would like to thank Dr Sun Weimin for improving the manuscript.
\end{acknowledgments}

\appendix
\section{Rotational Invariance}
In this Appendix we shall give a proof for the statistically rotational invariance of $T^{s}(\vec{n}^{'}) T^{s}(\vec{n})$ and rotational invariance of $F(\vec{\Delta})$.

From Eq.(\ref{talm}), $< T^{s}(\vec{n}^{'}) T^{s}(\vec{n}) >$ can be expressed as
\begin{eqnarray}
&& < T^{s}(\vec{n}^{'}) T^{s}(\vec{n}) >  \nonumber\\
& = & \int d^{3}\vec{k}^{'} \int d^{3}\vec{k} \int^{\eta_{0}}_{0} d\eta^{'} S_{T}^{s}(\eta,k^{'}) e^{ix^{'}\hat{k}^{'} \cdot \hat{n}^{'}} \nonumber\\
&& \int^{\eta_{0}}_{0} d\eta S_{T}^{s}(\eta,k) e^{-ix\hat{k} \cdot \hat{n}} < \xi(\vec{k}) \xi^{*}(\vec{k}^{'}) >   \nonumber\\
& = & \int d^{3}\vec{k}^{'} \int d^{3}\vec{k} \int^{\eta_{0}}_{0} d\eta^{'} S_{T}^{s}(\eta,k^{'}) e^{ix^{'}\hat{k}^{'} \cdot \hat{n}^{'}}   \nonumber\\
&& \int^{\eta_{0}}_{0} d\eta S_{T}^{s}(\eta,k) e^{-ix\hat{k} \cdot \hat{n}} f(k,k^{'},|\vec{k}-\vec{k}^{'}|).
\end{eqnarray}
Then the two point ensemble statistical average after rotation $R$ is
\begin{eqnarray}
&& < T^{s}(R\vec{n}^{'}) T^{s}(R\vec{n}) >  \nonumber\\
& = & \int d^{3}\vec{k}^{'} \int d^{3}\vec{k} \int^{\eta_{0}}_{0} d\eta^{'} S_{T}^{s}(\eta,k^{'}) e^{ix^{'}\hat{k}^{'} \cdot R\hat{n}^{'}}  \nonumber\\
&& \int^{\eta_{0}}_{0} d\eta S_{T}^{s}(\eta,k) e^{-ix\hat{k} \cdot R\hat{n}} f(k,k^{'},|\vec{k}-\vec{k}^{'}|).
\end{eqnarray}
Replace the dummy variables $\vec{k},\vec{k}{'}$ by $R\vec{k},R\vec{k}{'}$ and note that $d^{3}(R\vec{k}) = d^{3}\vec{k}$, $|R\vec{k}| = |\vec{k}|$, $d^{3}(R\vec{k}^{'}) = d^{3}\vec{k}^{'}$, $|R\vec{k}^{'}| = |\vec{k}^{'}|$ and $|R(\vec{k} - \vec{k}^{'})| = |\vec{k} - \vec{k}^{'}|$, one obtains
\begin{eqnarray}
&& < T^{s}(R\vec{n}^{'}) T^{s}(R\vec{n}) >   \nonumber\\
& = & \int d^{3}\vec{k}^{'} \int d^{3}\vec{k} \int^{\eta_{0}}_{0} d\eta^{'} S_{T}^{s}(\eta,k^{'}) e^{ix^{'} R\hat{k}^{'} \cdot R\hat{n}^{'}}  \nonumber\\
&& \int^{\eta_{0}}_{0} d\eta S_{T}^{s}(\eta,k) e^{-ix R\hat{k} \cdot R\hat{n}} f(k,k^{'},|\vec{k}-\vec{k}^{'}|).
\end{eqnarray}
Since $R\hat{k}^{'} \cdot R\hat{n}^{'} = \hat{k}^{'} \cdot \hat{n}^{'}$ and $R\hat{k} \cdot R\hat{n} = \hat{k} \cdot \hat{n}$, one finally obtain
\begin{eqnarray}
< T^{s}(R \vec{n}^{'}) T^{s}(R \vec{n}) > = < T^{s}(\vec{n}^{'}) T^{s}(\vec{n}) >.
\end{eqnarray}
The same reasoning can be applied to $F(\vec{\Delta})$. From Eq.(\ref{fd}), one obtains
\begin{eqnarray}
F(R \vec{\Delta}) & = & \dfrac{(4\pi)^{2}}{2l+1} \sum_{m}  \int d^{3}\vec{k}^{'} d^{3}\vec{k}  G_{l}(k^{'}) G_{l}(k) Y_{lm}(\hat{k}^{'}) \nonumber\\
&& Y_{lm}^{*}(\hat{k}) f(k,k^{'},|\vec{\Delta}|) \delta^{3} (\vec{k}-\vec{k}^{'} - R\vec{\Delta}).
\end{eqnarray}
Replace the dummy variables $\vec{k},\vec{k}{'}$ by $R\vec{k},R\vec{k}{'}$ as before and noted that $d^{3}(R\vec{k}) = d^{3}\vec{k}$, $|R\vec{k}| = |\vec{k}|$, $d^{3}(R\vec{k}^{'}) = d^{3}\vec{k}^{'}$, $|R\vec{k}^{'}| = |\vec{k}^{'}|$, $|R\vec{\Delta}| = |\vec{\Delta}|$, $Y_{lm}(R\vec{k}^{'}) = \sum_{m^{'}} D^{l}_{mm^{'}}(R) Y_{lm}(\vec{k}^{'})$, $Y^{*}_{lm}(R\vec{k}) = \sum_{m^{'}} D^{l*}_{mm^{'}}(R) Y^{*}_{lm}(\vec{k})$ and $\delta^{3} [R (\vec{k}-\vec{k}^{'} - \vec{\Delta}) ] = \dfrac{\delta^{3} (\vec{k}-\vec{k}^{'} - \vec{\Delta})}{|R|} =\delta^{3} (\vec{k}-\vec{k}^{'} - \vec{\Delta})$, one gets
\begin{eqnarray}
F(R \vec{\Delta}) = F(\vec{\Delta}).
\end{eqnarray}.


\end{document}